\documentclass[12 pt,eps]{article}
\usepackage[greek,english]{babel}
\usepackage[dvips]{graphicx}
\usepackage{epsfig}
\usepackage{fancyheadings}
\usepackage[greek,english]{babel}
\usepackage{graphicx}
\usepackage{amsmath}
\usepackage{theorem}
\usepackage{amssymb}
\usepackage{latexsym}

\catcode`\@=11
\def\Bbb{\ifmmode\let\next\Bbb@\else
\def\next{\errmessage{Use \string\Bbb\space only in math mode}}\fi\next}
\def\Bbb@#1{{\Bbb@@{#1}}}
\def\Bbb@@#1{\fam\msbfam#1}

\else
\fi \topmargin -30pt

\ifcase \@ptsize
\or 
\or 
\fi

\def \sg{sine-Gordon }
\def \csg{complex sine-Gordon }

\def \sm{$S$-matrix }

\def \d{\partial}
\def \db{\bar{\partial}}

\def \dt {\partial_{0}}
\def \dx {\partial_{1}}

\def \b{\beta}

\def \t{\theta}

\def \us {u^{*}}

\def \u {u^{*}}

\def \be{\begin{eqnarray}}
\def \ee{\end{eqnarray}}

\def \x{\chi}

\begin{document}

\title {Quantum complex sine-Gordon model on a half line}
\author{P. Bowcock, G. Tzamtzis}
\maketitle

\bigskip
\centerline{Department of Mathematical Sciences}
\centerline{University of Durham} \centerline{Durham DH1 3LE,
England}
\medskip
\vskip 4pc

\begin{abstract}
 In this paper, we examine the quantum complex sine-Gordon model on
 a half line. We obtain the quantum spectrum of boundary bound states  using  the
the semi-classical method of Dashen, Hasslacher and Neveu. The results are compared and found to agree with the bootstrap programme. A particle/soliton reflection factor is conjectured, which is consistent with unitary, crossing and our semi-classical results.

\end{abstract}

\section{Introduction}
\

 The \csg model \cite{Lund:1976ze,Pohlmeyer:1975nb}, appears as the
 simplest case in a family of homogeneous generalisations for the
 \sg model\cite{Fernandez-Pousa:1997hi,Fernandez-Pousa:1998iu,
  Castro-Alvaredo:1999em,Dorey:2002sc}.
  It is an integrable $(1+1)$ dimensional field theory
 with an internal $U(1)$ degree of freedom. It is described by a
 Lagrangian of the form
 \be \label{eq:CSGlagr2}
 {\cal L}_{CSG}  =  \frac{1}{2}\frac { {\d u} {\db \us} +{\db u}
 {\d \us}}{1 - \xi^2 u u^{*}}- m^2 u u^{*} \  ,
 \ee
 where $u$ is a complex field and $\xi$ is a real coupling
 constant. The model accepts soliton \cite{Lund:1976ze,Getmanov:1977hk}
 and breather solutions\cite{Tzamtzis:2002} which are non topological in nature. The
 \sg model is recovered only in the chargeless limit, in which solutions
 become topological objects.

 In a previous paper \cite{Tzamtzis:2002} we studied the classical \csg
 theory in the bulk and on a half line.  
 The behaviour of various solutions and  their interaction with the boundary was analysed. Through the construction of
 conserved currents we managed to obtain a general  boundary
 condition that preserves integrability
 \be\label{eq:CSGintbc}
 \dx u = -C u \sqrt{1- u \u} &,&  \dx \u = -C \u \sqrt{1- u \u} \
 ,
 \ee
 where  both expressions are evaluated at $x=0$ and $C$ is a real parameter.
 The full Lagrangian on a half line is
 \be \label{eq:CSGlagr2}
 {\cal L}_{CSG}  = \int^0_-\infty \,dx\left (\frac { {\dt u} {\dt \us} -{\dx u}
 {\dx \us}}{1 - \xi^2 u u^{*}}- m^2 u u^{*} \right )+ \left[2 C \sqrt{1-u
 \u} \right]_{x=0} \ .
 \ee
In this paper,
 we shall aim to extend our study of  the model on the half-line to the quantum regime. 
 
 The outline of this paper is as follows: in the second section we
 review the most important results for the quantum \csg theory in
 the bulk.

 In the third section we shall use the  semi-classical
 stationary-phase approach
 which is the field theory analogue of the WKB method of quantum
 mechanics. A generalised version of  Bohr-Sommerfeld
 quantisation, following the method of Dashen, Hasslacher and Neveu  \cite{Dashen:1975hd,Dashen:1974ci} will be used to obtain the semi-classical spectrum of boundary states
 and their first order corrections. As we shall see, these
 corrections induce  a finite renormalisation of both the bulk and boundary coupling
 constants. The former is consistent with results from the bulk case.

 Following that,
 the bootstrap programme devised by Ghoshal and
 Zamolodchikov  \cite{Ghoshal:1994tm} will be used to construct
 the quantum reflection factors of particles and solitons scattering off the boundary. The results of 
 Delius and Gandenberger \cite{Delius:1999cs} are adapted to find to postulate a reflection factor
 which is consistent with the bulk scattering matrix found by Dorey and Hollowood \cite{Dorey:1995mg}, crossing, unitarity and semi-classical results.
 In particular, the use of carefully chosen CDD factors furnishes these factors with poles consistent with  the existence of the boundary-bound states, found in the previous section. That this can be done consistently provides some evidence in favour of our conjecture.

 The paper finishes with a few general remarks about the results of
 the quantum case including a brief discussion about the poles in
 the bootstrap method and the necessary conditions in order for them
 to lie within the physical strip.

\section{Quantum \csg theory in the bulk}
\

 The \csg model as a quantum field theory in the bulk was
 studied relatively soon after the model's introduction. The classical
 treatment which showed the theory to be completely integrable
 and to possess soliton solutions carrying  a $U(1)$ charge, prompted
 researchers to look into the quantum case in the hope that the
 nice features of the model persisted in this limit too.

 The investigation of the quantum case began with the work of
 de Vega and Maillet  \cite{deVega:1981ka} in which they showed that the \sm
 is factorisable at tree level. The model remains integrable and
 continues to accept soliton solutions in the quantum case.
 Provided that a specific counterterm which  depends on the field
 is added to the Lagrangian, the \sm is also factorisable at one-loop level.
 In their following paper \cite{deVega:1983sh} they used
 the inverse scattering method to obtain the classical two-soliton solution
 and the spectrum of states using the  semi-classical methods
 by Dashen, Hasslacher and Neveu \cite{Dashen:1975hd,Dashen:1974ci}.

 The two-loop order case was studied by Bonneau in \cite{Bonneau:1983hx} who,
 continuing down the path of de Vega and Maillet, showed that the theory
 is non-renormalisable unless a finite number of counterterms
 (quantum corrections) are added.

 After a gap of almost ten years the quantum \csg case was
 revisited by Dorey and Hollowood \cite{Dorey:1995mg} in the light of the
 theory emerging as a gauged WZW model \cite{Bakas:1994xh}. With the
 semi-classical results of de Vega and Maillet as a guide, they
 proposed an exact \sm based on the demands of the bootstrap programme.
 We begin our review with the general form of a single soliton solution
 \be \label{eq:Usoliton}
 u = \frac{\cos(a) \exp {\left[ i m \sin(a) ( \cosh(\t) t - \sinh(\t) x
 )\right]}}{  \cosh \left( m \cos(a)( \cosh{(\t)}( x -x_0)
 \sinh{(\t)} t)\right) } ,
 \ee
 where $a$ is a real parameter associated with the $U(1)$ charge,
 and $\t$ the rapidity of the solution. From the expressions for the energy
 and charge
 \be \label{eq:bulkenergy}
 {E} = \int \frac{|\dt u|^2 +|\dx u|^2}{1 - u \us}
 + m^2 u \us &\ , \ &
 Q = i \int dx \frac{\u \dt u - u \dt \u}{1 -  u \u } \  ,
 \ee
 we easily obtain for the single soliton case
 \be \label{eq:mq}
 E=\frac{4 m}{\xi^2} \cos(a)\cosh(\t) & \ , \ &  Q = \frac{4}{\xi^2}
 \left(\mbox{sign[a]}\frac{\pi}{2} - a \right) \ .
 \ee
 In their paper Dorey and Hollowood argued that it is necessary to
 only consider specific values for the coupling constant. Specifically
 they argued that the only acceptable values are
 \be
 \xi^2 = \frac{ 4 \pi }{k} &,& k>1 \ ,
 \ee
 where $k$ is an integer. This agrees with the WZW interpretation of the
 theory where $k$ corresponds to the level of the $SU(2)/U(1)$ coset
 model. With this constraint their proposed \sm reproduces the
 semi-classical spectrum of states derived by de Vega and Maillet.
 In addition they proposed that the charge is conserved if it is
 defined modulo $k$. In the following,
 we shall use $k$ instead of the coupling constant $\xi$ for reasons
 of simplicity.

\begin{figure}[htb]
 \begin{center}
 \fbox{
 \includegraphics[width=.75\textwidth,height=.5\textwidth]{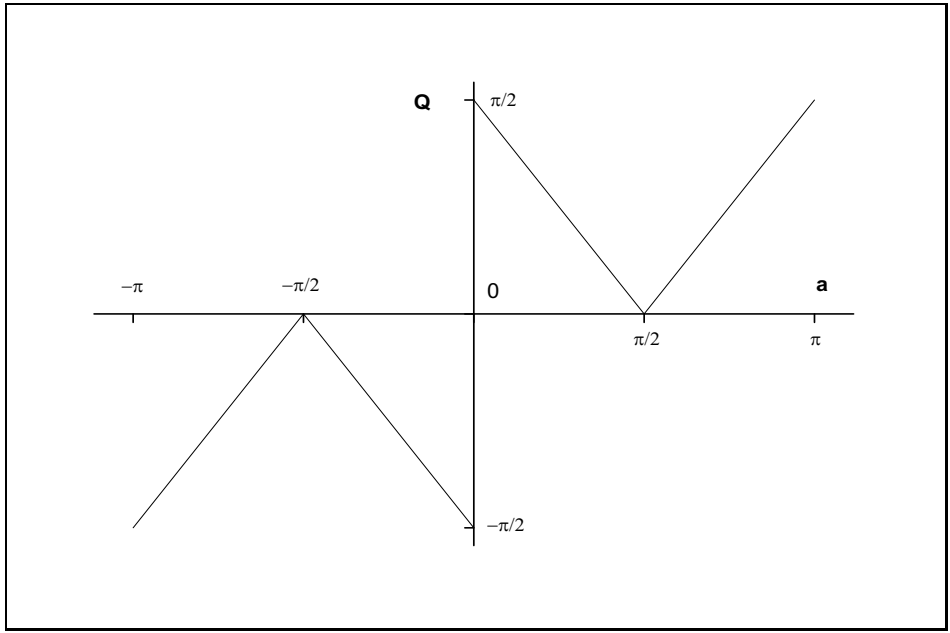}}
 \end{center}
 \caption{Soliton charge Q($a$)}\label{fig:Qsoliton}
 \end{figure}

 The quantum spectrum can be found by using the Bohr-Sommerfeld
 quantisation rule
 \be
 S(u) + E(u) \tau = 2 \pi n \ &,& n \ \epsilon \ \mathbb{Z} \ ,
 \ee
 where $S$ is the action functional, $E(u)$ the energy and
 $\tau$ the period of the solution $u$. No topological
 distinction exists between the vacuum and the soliton sector,
and thelowest-charge soliton may be regarded as the the basic particle of the CSG
 theory. The static one-soliton solution is
 \be \label{eq:staticsoliton}
 u_{static} =  \frac{\cos(a) \exp \left( i m \sin(a) { t }\right)}
 {\cosh \left({ m \cos(a) { (x-x_0)
 }}\right)} \  .
 \ee
 This solution is not time-independent. It does not translate in the $x$
 direction but oscillates in a breather-like fashion. As pointed out
 by Ventura and Marques \cite{Ventura:1976vi}, and Montonen
 \cite{Montonen:1976yk}, the Bohr-Sommerfeld quantisation is equal to
 charge quantisation for scalar field theories enjoying  a global
 $U(1)$ symmetry. For the CSG case the only time dependence for the
 soliton  solution in the rest frame is restricted to the phase i.e.
 $u \u$  does not  depend on time. It is easy to show that for the
 static one-soliton
 \be
 S(u) + E(u) \tau = 2 \pi Q \ ,
 \ee
 which in turn implies
 \be
 Q = n .
 \ee
 This corresponds
 to a tower of states with ever increasing charge. However in
 the classical case the charge is a periodic function. In the theory
 $a$ appears only in trigonometric forms, and therefore should be
 considered as an angle variable. This indicates that the formula
 for $Q$ which has previously appeared in the literature is only true
 for a certain region, namely $-\frac{\pi}{2} \leq a \leq \frac{\pi}{2}$.
 If one plots $Q$ as a function of $a$ the result is a periodic
 pattern (fig. \ref{fig:Qsoliton}). To avoid confusion we shall
 consider $a$ to lay in the region $0\leq a \leq \frac{\pi}{2}$,
 unless stated otherwise.

 As suggested by Dorey and Hollowood the charge should be
 defined modulo $k$ which leads to a finite spectrum depending on
 $k$
 \be
 Q =& \pm 1 , \pm 2 , ...,\pm \frac{k-1}{2}  &\ \ \ \mbox{for} \ k \
 \ \mbox {odd} \ ,
 \\
 Q =& \pm 1 , \pm 2 , ...,\pm \frac{k-2}{2},\frac{k}{2}  &\ \ \
 \mbox{for} \ k \ \ \mbox{even} \  \ .
 \ee
 From the classical expression for the charge (\ref{eq:bulkenergy}) we have
 \be
 Q = \frac{k}{\pi} (\frac{\pi}{2} - a) \ .
 \ee
 The quantisation of the charge is in fact the quantisation of  $a$
 \be
 2 \pi Q  = 2 \pi n \Leftrightarrow a = \frac{\pi}{k}\left( \frac{k}{2}
 -n \right) \ .
 \ee
 Through the quantisation of the charge parameter, the energy spectrum
 \be
 M = \frac{k}{\pi} m \cos (a) \ ,
 \ee
 is also obtained
 \be
 M (Q) = \frac{ k m }{\pi} \left| \ \sin \left(\frac{ \pi Q }{k}\right)
 \right| \ .
 \ee
 The Bohr-Sommerfeld quantisation provides us only with the quantum
 spectrum up to leading order. According to de Vega and Maillet
 the next order corrections is achieved by a simple
 renormalisation of the coupling constant
 \be
 \xi^2 \rightarrow \xi_{R}^2 = \frac{\xi^2} {1-\xi^2/4 \pi} \ .
 \ee
 or equivalently
 \be \label{renormalisation}
 k \rightarrow k_R = k -1  \ .
 \ee
At this level an  \sm may be written down that reproduces
 the semi-classical spectrum to leading order. In their paper
 Dorey and Hollowood first presented a minimal choice for the
 meson-soliton scattering matrix
 \be
 S_{1Q} = F_{Q-1}(\t)F_{Q+1}(\t) \ .
 \ee
 which reproduced the semi-classical behaviour and agreed
 with the results of deVega and Maillet. Moreover, through this
 they confirmed that the meson in the CSG theory can be identified
 with the $Q=1$ soliton. The function $F_Q(\t)$ is defined as
 \be \label{eq:ffactor}
 F_Q(\t) = \frac{ \sinh \left( \frac{\t}{2} + \frac{ i \pi}{2 k} Q \right)}
        { \sinh \left( \frac{\t}{2} - \frac{ i \pi}{2 k} Q\right)} \
        .
 \ee
 $S$-matrices constructed from products of the $F_Q(\t)$
 automatically satisfy unitarity and analyticity constraints. With the above
 result as a starting point they proposed the following \sm for
 arbitrary charge
 \be\label{eq:smatrix}
  S_{Q_1Q_2} =  F_{Q_1-Q_2} \left[ \prod_{n=1}^{Q_2-1}  F_{Q_1 - Q_2
  + 2 n  }\right]^2 F_{Q_1+Q_2} ,
 \ee
 which satisfies all the familiar restrictions and has the correct
 pole structure. In addition they pointed out that this is
 exactly the minimal \sm associated with the Lie algebra $a_{k-1}$
 and conjectured that for the specific choices of the coupling this
 \sm should be exact.

 With the general form of the \sm the review of the
 bulk case is concluded. More details on the above
 results may be found in the relevant papers. Nevertheless, this
 presentation contain all the ingredients needed to examine the
 half-line case.

 \section{Semi-classical quantisation}
 \

 In this section we shall attempt to build the complete spectrum of
 quantum boundary-bound states for the \csg model on a half line. The fact that
 the model possesses exact periodic solutions bound to the boundary makes the
 stationary-phase method a suitable candidate for their
 quantisation. The method is based on the work of Dashen, Hasslacher
 and Neveu \cite{Dashen:1975hd,Dashen:1974ci} for the semi-classical
 quantisation of the \sg model. In this case it appears as a
 generalised version of the Bohr-Sommerfeld quantisation rule
 \be
 S_{qu} - T \frac{\d S_{qu}}{\d T} = 2 \pi n \ ,
 \ee
 where $S_{qu} = S_{cl} - \Delta$. The parameter $\Delta$ is
 related to the stability angles of the stationary phase approach
 and produces the first order quantum corrections to the classical
 action.  We shall use the ordinary quantisation condition for the classical
 action $S_{cl}$  and then we shall calculate the quantum
 corrections factor
  \be \label{eq:quantumcorrection}
 \Psi = \left( \Delta - T \frac{\d \Delta}{\d T} \right) \ .
 \ee
 
 Before embarking on quantisation, we will recall a few of the classical 
 results about the boundary spectrum from \cite{Tzamtzis:2002}. The energy of
 the half-line theory is given by
 \be
 E=\int_{-\infty}^0 \left [\frac{|\dt u|^2 +|\dx u|^2}{1 - u \us}
 + m^2 u \right ] -2C\sqrt{1-uu^*}
 \ee
 For a given value of $C$, the lowest energy configuration corresponds to the 
 vacuum $u=0$. Linearising the equations of motion and the boundary conditions (\ref{eq:CSGintbc}) one calculates the particle reflection coefficient can be found to be
 \be\label{eq:classicalref}
 K(k)=\frac{ik+C}{ik-C}
 \ee
 where $k$ is the momentum of the particle. For $C<-m$, one can show that the pole corresponds to a normalisable but in time  exponentially growing perturbative mode showing that  the vacuum is an unstable configuration. For $C$ in this range, the boundary 
 can emit a real soliton which effectively swaps the sign of $C$ so that ultimately the classical field configuration returns to $u=0$, but now with $C>m$.
 For $|C|\leq m$, it is possible to construct a classical excited boundary state by
 taking $u=u_{cl}$ to be a static soliton solution of the form in (\ref{eq:staticsoliton}), provided
 that the parameters are chosen to obey the equation
 \be\label{eq:solitonbc}
 C=\pm\frac{m}{\sqrt{1+\tan^2(a)\coth^2(m\cos(a)x_0)}}
 \ee
 There are other more complicated boundary states based on multi-soliton solutions
 such as breathers \cite{Tzamtzis:2002}, but a thorough analysis of these is beyond the scope of the present study.
 
 The boundary term of (\ref{eq:CSGlagr2}) preserves the $U(1)$
 charge as it only depends on mod($u$). The theory remains $U(1)$
 invariant and therefore the Bohr-Sommerfeld quantisation is
 equivalent to charge quantisation on the half line case as it was
 for the bulk. The quantisation condition for the classical action reads
 \be
 S_{cl}[u_{cl}] + E_{cl}[u_{cl}] T = 2\pi Q = 2 \pi n \ .
 \ee
 with $u_{cl}$ the static one soliton solution of (\ref{eq:staticsoliton})
 with parameters chosen to satisfy (\ref{eq:solitonbc}).
 It is a periodic solution with period
 $T$ exhibiting a breather-like
 behaviour. It provides the perfect starting point as the simplest
 boundary bound state of the classical theory.  The position of
 the centre of mass will determine the charge of the bound state
 as a fraction of the  soliton charge in the bulk.
 We can calculate the charge of the static soliton through the
 expression
 \be\label{eq:Q2}
 Q =- i \int_{-\infty}^{0} dx \frac{\u \dt u - u \dt \u}{1 -  u \u } \
 .
 \ee
 Note that this charge is conserved without the need to add a term at the boundary.
 This time however the integration takes place on the half line
 and finally yields
 \be\label{eq:chargeboundary}
 Q = \frac{k }{2\pi} \left( \pi - b - a \right) \ .
 \ee
 The charge now depends on the boundary parameter $C=m\cos(b)$,
 which enters the calculation through the position of the centre of mass.
 As a check that this formula is correct, we can set
 $b=\frac{\pi}{2}$ which implies that we have the Neumann boundary condition $\partial_x u=0$,
 thus placing the soliton at exactly $x=0$. The charge is then exactly
 half of the equivalent charge of the soliton in the bulk (\ref{eq:mq})
 as expected. The quantisation condition now reads
 \be \label{eq:quant1}
 Q = \frac{k }{2\pi} \left( \pi - b - a \right) = n \ ,
 \ee
 or in terms of the charge parameter
 \be
 a =   \pi - b -\frac{2 \pi n}{k} \ .
 \ee
 The quantisation of $a$ provides us with the
 first approximation of the semi-classical spectrum
 \be
 E_n =  \frac{k m}{2 \pi} \cos (a) = \frac{k m}{2 \pi}
      \cos(\pi - b -\frac{2 \pi n}{k}) \ .
 \ee
 Having a general formula for the energy, it is useful to
 calculate the energy difference between two adjacent states. We
 shall use this in the following section where we shall
 compare it with the corresponding bootstrap result. For 
 reasons of simplicity, we shall assume
 that the values of the parameters are such that the cosine is
 positive for both states as is their difference so that we can ignore
 any modulus appearing. The
 energy difference is then written
 \be
 E_{n+1} - E_n =  \frac{k m}{2 \pi} \left(\cos(\pi - b -
 \frac{2 \pi (n+1)}{k})- \cos(\pi - b -\frac{2 \pi n}{k}) \right)\ ,
 \ee
 which after some manipulation simplifies to
 \be\label{eq:energydifference}
 E_{n+1} - E_n =  \frac{k m}{\pi}\cos\left(\frac{\pi}{2}-\frac{\pi}{k}
 \right) \cos\left(\frac{\pi(2n+1)}{k} -\frac{\pi}{2} + b)\right)\ .
 \ee
 We shall return to this result at the end of the next section,
 after we have obtained a relevant expression from the bootstrap
 programme.

Having obtained the leading order quantisation of the boundary spectrum,
we calculate the first-order correction to this.
 The step consists of calculating the quantity $\Delta$
 \be \label{eq:Delta}
 \Delta = - S_{ct} + \sum_{i=0}^{\infty} \frac{1}{2} \ \hbar
 \ v_i \ ,
 \ee
 where $v_i$ are the stability angles and $S_{ct}$
 suitably selected counter terms to cancel any arising infinities.
 We shall first calculate the sum over the stability
 angles, while the counter terms will be introduced later.
 In order to obtain all possible states one
 has to consider the theory in a box. The theory is already
 bounded on the right from  the original boundary, therefore
 another boundary should be introduced restricting the system
 in the finite volume $[-L,0]$. Using periodic boundary conditions
 will force all energy levels to become discrete. We can
 afterwards take the limit $L\rightarrow -\infty$ to recover the
 original system. The most appropriate choice would be Dirichlet
 boundary conditions $u=0$, which reproduce the correct behaviour
 of $u$ at infinity.

 The stability angles are obtained by solving the  linearised stability
 equation  about  a given classical solution. In our
 case we perturb around the static one-soliton solution. Once the
 solutions $\x(x,t)$ for the stability equation are found then the stability
 angles can be calculated from periodicity demands
 \be\label{eq:stabilityangles}
 \x_i(x,t+T) = e^{i v_i} \x_i(x,t) \ ,
 \ee
 Instead of  solving directly the stability equation we can follow the method
 of Corrigan and Delius to calculate the sum of the stability
 angles through the reflection factors. We begin with the classical
 two-soliton solution of the CSG model which satisfies the
 stability equation.  By fixing  the free parameters we can make one of the
 solitons $S_1$ static by taking $\delta_1=1$ and the other $S_2$ very
 small by taking the charge parameter $a_2$ close to zero. Effectively
 we are left with a small perturbation around a static one-soliton
 background. This is exactly the same method that  Dashen, Hasslacher
 and Neveu followed to calculate the stability angles of the sine-Gordon
 model. At infinity the static soliton is practically zero whilst the
 perturbation appears as plane waves
 \be
 \x(x,t) = e^{-i\omega t} ( e^{i k_s x} + R_s e^{-i k_s
 x})     \ .
 \ee
 The reflection factor is found to be
 \be
 R_s  = - \frac{(\cos(b -i \t))}{(\cos(b + i \t))}
         \ \frac{(\sin(a+i\t)-1)}{(\sin(a-i\t)-1)}
  \ ,
 \ee
 where we have set $m=1$ without loss of generality. The rapidity $\t$
 which has been used is related to the momentum through $k=\sinh(\t)$.
 The parameters $a$ and $b$ are the familiar charge and boundary
 parameters.
 From the above expression for $\x(x,t)$ and (\ref{eq:stabilityangles})
 we can  substitute
 \be
 \Delta = \frac{1}{2} \sum_i v_i = \frac{T}{2} \sum_i \omega_i \ ,
 \ee
 where $\omega_i^2 = k_i^2 + 1$ and the counter terms have been
 neglected. Placing the system in a box allows for discrete values
 of $k$ which should not be confused with the coupling constant.
 From $\Delta$ we must now subtract the vacuum contributions
 \be\label{eq:Delta2}
 \Delta = \frac{T}{2} \sum_i \sqrt{{k_{s,i}^2 + 1}} -
 \sqrt{k_{0,i}^2
 + 1} \ .
 \ee
 From the Dirichlet boundary conditions we can obtain the
 following equation relating the discrete momenta with the
 reflection factors
 \be
 e^{- 2 i k_s L} =   \frac{(\cos(b-i \t))}{(\cos(b + i \t))}
         \ \frac{(\sin(a+i\t)-1)}{(\sin(a-i\t)-1)} \ .
 \ee
 A similar equation exists for the fluctuations around the vacuum
 \be
 e^{- 2 i k_0 L}= - \frac{i \sinh(\t) + \cos(b)}{i \sinh(\t) -
 \cos(b)} \ .
 \ee
 Using the same argument as Corrigan and Delius, we can define a
 function $\kappa(k_0)$ such that for large $k$ we may write
 \be \label{eq:ks}
 k_ s = k_0 + \frac{\kappa(k_0)}{L} \ ,
 \ee
 where the index $i$ has been suppressed. This is possible since
 in the limit $\t \rightarrow + \infty$ and taking into account  the
 general quantisation condition of (\ref{eq:quant1}), both reflection
 factors $R_s$ and $R_0$ are equal.  Through the difference
  $k_s-k_0$ a function $\kappa(\t)$ may be defined using the
 ratio of the corresponding reflection factors
 \be
 e^{-2i \kappa(\t)} = - \frac{(\sin(a+i\t)-1)}{(\sin(a-i\t)-1)} \
                 \frac{(\sin(b+i\t)-1)}{(\sin(b-i\t)+1)} \ .
 \ee
 We can now calculate $\Delta$ in terms of $\kappa$. We can substitute
 (\ref{eq:ks}) in (\ref{eq:Delta2}) and then expand the expression
 in terms  of $L$. Keeping only the leading term of the expansion
 we end up with
 \be
 \Delta \sim \frac{T}{2L} \sum_i \frac{k_i^{(0)}\kappa(k_i^{(0)})}
  {\sqrt{(k_i^{(0)})^2+1}} \ ,
 \ee
 which in the limit $L\rightarrow +\infty$ can be substitute
 with the integral form
 \be
 \Delta =\frac{T}{2\pi} \int_0^{=\infty} dk \frac{k}{\sqrt{k^2+1}}\ \kappa(k)
 \ .
 \ee
 The calculation is greatly simplified if we change variables to
 $\t$
 \
 \be
 \Delta = \frac{T}{2} \int_0^{+\infty} d\t \sinh(\t) \kappa(\t) \ .
 \ee
 The integral is divergent but we can introduce specially chosen
 counter-terms to obtain a finite result. We begin with an
 integration by parts
 \be \label{eq:delta1}
 \Delta = \frac{T}{2\pi} \left(\left[ \kappa \cosh(\t)
 \right]_0^{+\infty} - \int_0^{+\infty} \cosh(\t)
 \frac{d\kappa}{d \t} \right) \ .
 \ee
 The first term is not divergent since the function $\kappa$
 approaches zero (through the quantisation condition) as $\t$
 goes to infinity. In this limit the combination $\kappa(\t)
 \cosh(\t)$ is zero. In addition, $\kappa(0) = 0$ so the first
 term in (\ref{eq:delta1}) vanishes. The second term is however
 divergent and counter terms  have to be introduced to cancel
 infinities. The latter appear in the same fashion as the
 logarithmic divergencies in the bulk which are tackled through
 normal ordering. With some straightforward manipulation the
 derivative term yields
 \be
 \frac{d\kappa}{d \t} =-{\frac {\cos (b) }{\cosh ({
 \t})- \sin (b) }}+{\frac {\cos (a) }{ \cosh ( {\t} )-\sin ( a ) }} \ .
 \ee
 The two terms are almost identical and both divergent. A logical
 choice of counter-terms seems to be
 \be
   -{\frac {\cos ( b ) }{\cosh ( \t )+1 }}+{
 \frac {\cos ( a ) }{\cosh ( \t )+1 }} \ ,
 \ee
 where the first removes the divergence associated with the
 boundary and the second with the one in the bulk. The complete
 expression to be calculated is now
 \be
 \Delta &=&  - \int_0^{+\infty}d\t \cosh(\t) \nonumber \\
  & \ & \left(
  {\frac {\cos (a) }{ \cosh ( {\t} )-\sin ( a ) }}
  - \frac {\cos ( a ) }{\cosh ( \t ) + 1 }
  -{\frac {\cos (b) }{\cosh ({ \t})- \sin (b) }}
  +{\frac {\cos ( b ) }{\cosh ( \t )+1 }}
  \right) , \nonumber
 \ee
 which finally yields
 \be
 \Delta = \frac{T}{2 \pi} \left( -\cos (a) +\cos (b) + b \sin (b)
  +\frac{\pi}{2} \,\sin (b) - a \sin (a) - \frac{\pi}{2} \sin ( a )
  \right) \ .
 \ee
 Having determined the form of $\Delta$ we are now in a position to
 calculate  the corrections to the classical Bohr-Sommerfeld rule.
 The expression of $\Delta$ depends on the period $T$ through the
 charge parameter $a$
 \be
 \sin(a) = \frac{2 \pi}{T} \ .
 \ee
 The correction term  of (\ref{eq:quantumcorrection}) is easily
 calculated
 \be
 \Delta - T \frac{\d \Delta}{\d T} =  - a -\frac{\pi}{2} \ .
 \ee
 With all the necessary parts calculated, the generalised
 Bohr-Sommerfeld quantisation condition finally reads
 \be
 k_r( \pi - b_r - a)  = 2 \pi n \ ,
 \ee
 where
 \be
 k_r = k-1 &,& b_r = \frac{ k b - \frac{3 \pi}{2}}{k-1} \ .
 \ee
 The form of the new quantisation condition is exactly the same as
 the first approximation of (\ref{eq:quant1}), only with a
 redefinition for the boundary and coupling constants.
 The shift in the coupling constant is to be expected.
 In the bulk case the first order corrections amount to a simple
 shift in $k$ (\ref{renormalisation}) as pointed by de Vega
 and Maillet \cite{deVega:1994pm}.
 Our result is consistent with this. 
 
 In addition to $k$, the boundary constant has to be renormalised
 as well. It is not clear why this renormalisation is needed or whether
 it should appear at all. In a  paper examining the closely
 related $a_{k-1}$ theory, Penati and Zanon \cite{Penati:1995bs}
 argued that renormalisation of boundary parameters has to be introduced
 in certain models to ensure integrability at the quantum level. The
 renormalisation of the coupling constant and its significance remains
 one of the open questions for the quantum CSG theory.

  \section{The Bootstrap Method}
 \

 In the previous section we constructed the quantum spectrum
 using the semi-classical stationary-phase method. Although the
 results are not exact, they provide us with an accurate picture
 of the set of states. The same spectrum can be obtained using the
 completely different approach of the bootstrap method.
 It is based on the pioneering work
 of Cherednik \cite{Cherednik:1984vs}, Ghoshal and Zamolodchickov
 \cite{Ghoshal:1994tm}, and Fring and K\"oberle \cite{Fring:1994mp}.
 This way we shall be able to compare and compliment the results to
 acquire an even more accurate  spectrum.
 The idea behind this method is to construct the reflection factors through the
 boundary bootstrap relation, and through
 the poles therein to identify boundary bound states. The process
 is analogue to the bulk case where the existence of bound states is
 indicated by poles found in the $S$-matrix.

 Nevertheless there are quite a lot of drawbacks in this process.
 First of all in order to begin  we need the  quantum reflection factor
 for the particle of the theory. This
 cannot be obtained through any consistent procedure. Although
 the principles of analyticity, unitarity and crossing symmetry
 along with the boundary Yang-Baxter equation restrict the form of
 the reflection factor enough, it is impossible to pin it down
 completely. It is therefore a matter of ``selecting" the correct
 reflection factor which should satisfy all constraints whilst
 introducing a pole corresponding to a boundary bound  state.

 After carefully deciding on the correct reflection factor for the
 particle, we are faced with a second problem. Although we are
 only concerned with poles in the reflection factor which appear
 on the physical strip (which in the half line case is
 $\Im(\t) \ \epsilon \ [0,\pi/2]$), not all of them correspond
 to boundary bound states. It is quite difficult to explain the
 appearance of all the poles which lie within the physical strip
 and no work until now exists to offer a systematic treatment of
 poles encountered.

 Another difficult arises with the determination of the CDD
 factors in the reflection matrix. The restrictions imposed on the
 reflection factor may determine it up to a set of factors, quite
 like the CDD factors of the $S$-matrix. However the reflection
 CDD factors which are also restricted by the same constraints,
 introduce more poles which are also difficult to explain.

 Considering all the above we shall attempt to construct the
 quantum boundary-bound state spectrum for the CSG theory but our study
 will be superficial. It is not our objective to fully explain the
 quantum structure of the model but rather to verify crudely our
 semi-classical results. The detailed explanation of poles in the
 reflection factors, the determination of a general form for the
 reflection CDD factors and the comparison of the results with similar
 theories are quite fascinating problems but are
 beyond the scope of this paper. In the following section we shall
 present a suitable form for the  quantum reflection factor
 for the particle of the CSG model.
 \\

 \subsection{Quantum reflection factor for the CSG particle}
 \

 We begin our attempt to introduce a suitable reflection matrix
 $K_1$ for the CSG particle with the assumption that it should be
 made out of $F$ factors that were defined in (\ref{eq:ffactor}).
 It is a natural selection as both the boundary Yang-Baxter equation
 and the bootstrap equations relate reflection matrices with
 the \sm  which is expressed only in terms of such functions.
 There is another advantage to this choice. Unitarity, real
 analyticity and $2\pi i$-periodicity requirements are
 automatically satisfied if $K$ appears as a product of such
 factors. Some of the most important properties enjoyed by these
 functions include the following
 \be
 F_Q(\t)F_Q(-\t)= 1 \ , & \ F_{2Q} (2 \t) = - F_{Q}(\t)F_{Q+k}(\t)
 \ , & \ F_Q(\t+i\pi) = - F_{Q+k}(\t) \ . \nonumber
 \ee
 The first of the above is responsible for the fulfillment of
 the unitarity requirement. The remaining demonstrate basic
 transformations between rapidity and charge and will be used in
 the bootstrap and crossing symmetry equations.

 It should be noted that since there is no degeneracy in the
 spectrum we expect $K$ to be diagonal. This, in
 conjunction with a diagonal $S$-matrix, renders the boundary Yang-Baxter
 equation
 \be \label{eq:boundaryYangbaxter}
 K_a(\t_a)S_{ab}(\t_a+\t_b) K_b(\t_b) S_{ab}(\t_a-\t_b)=  S_{ab}(\t_a-\t_b)
 K_b(\t_b)  S_{ab} (\t_a+ \t_b )K_a (\t_a)\ , \nonumber
 \ee
 trivial. In their paper Dorey and Hollowood identified the CSG particle
 with the $Q=1$ soliton. In this context the first real constraint
 for $K$ comes from the crossing symmetry relation
 which for our case is written as
 \be \label{eq:boundcrossing}
 K_1(\t) K_{\bar{1}}(\t + i \pi ) = S_{1,1} ( 2 \t)  \ ,
 \ee
 where $K_{\bar{1}}$ is the reflection of the antiparticle and
 $S_{1,1}$ is the two particle scattering matrix. Dorey and
 Hollowood noted that the CSG \sm is identical to the minimal
 $a_{k-1}$ \sm  which in turn can be recovered from the  $a_{k-1}^{(1)}$
 Affine Toda field theory (ATFT) when the parts involving the coupling constant
 are omitted. It is therefore reasonable to
 build our reflection matrix based on the proposed form for the
 particle reflection matrix of the boundary $a_{k-1}^{(1)}$ ATFT theory.
 In their paper Delius and Gandenberger \cite{Delius:1999cs}
 present a general form for the particle reflection matrix of the
 $a_{k-1}^{(1)}$ ATFT. As in the
 \sm case the reflection matrix is a product of two parts, out
 of which only one depends on the coupling constant.
 Each part satisfies the bootstrap independently
 so we can recover a $K$-matrix for our model by simply ignoring
 the coupling dependent pieces. The block
 notation implies $(x)= F_x(\t)$ and shall be used
 henceforth in parallel with $F$. Ignoring the
 parts involving the coupling constant, the remaining factors
 \be \label{eq:reflectionbase}
 K_n =  \sum_{c=1}^n (c-1) (c-k) &,& n=1..(k-1) \ ,
 \ee
 constitute a complete set satisfying the crossing-symmetry condition
 (\ref{eq:boundcrossing}) as well as the reflection bootstrap equation
  \be\label{eq:reflectionbootstrap}
 K_c(\t_c) = K_a(\t_a) K_b(\t_b) S_{ab}(\t_a+\t_b) \ .
 \ee
 This is not unexpected as both theories
 share the same minimal \sm. This however creates a problem as $K$ does
 not contain any poles which can be related to boundary-bound states.
 This means that should any additional factors be added by hand to
 introduce the required poles, they should be added in such a way that
 they cancel between them in the crossing and  bootstrap
 relations. This in turn suggests that the new factors are nothing
 more than CDD factors for the reflection matrix. Placing the poles in
 the CDD factors simplifies the whole procedure of determining a suitable
 particle reflection matrix since they satisfy simpler relations.

 Real analyticity and unitarity conditions once again prompt us to
 construct our CDD factors out of  block functions, so that
 the former are satisfied automatically. In order
 for the crossing relation to be satisfied any block factor $(x)$
 should be  accompanied by the charge conjugate factor $(k-x)$.
 Delius and Gandenberger showed for such a combination the
 bootstrap closes.

\begin{figure}[hbt]
 \begin{center}
 \fbox{
 \includegraphics[width=.75\textwidth,height=.5\textwidth]
 {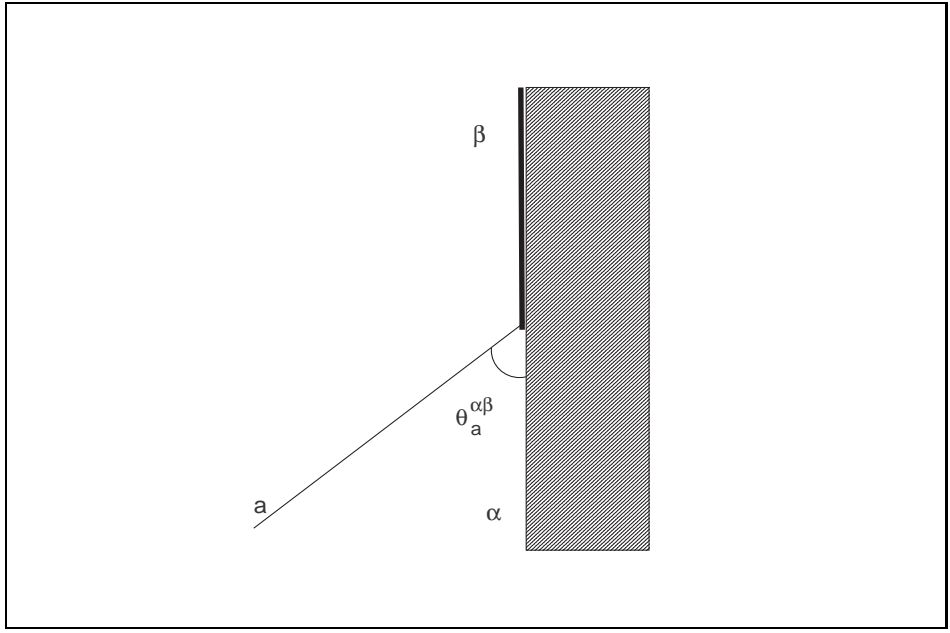}}
 \end{center}
 \caption{The boundary bound state}
 \label{fig:boundarybound}
 \end{figure}

 Now that we have a  consistent way of adding factors we need to find
 where the poles should appear. We begin with the simple formation
 of the boundary bound state described in (Fig. \ref{fig:boundarybound}).
 During the process both energy and charge are conserved. We begin
 with the charge conservation. Far  away from the boundary the soliton
 (particle) behaves as in the
 theory in the bulk. Its' charge is equal to the normal soliton
 charge $Q_1= \frac{k}{\pi}(\frac{\pi}{2}-a_1)$. After the
 formation of the bound state the charge $Q_2$ is given by the formula
 (\ref{eq:chargeboundary}). Equating these yields
 \be\label{eq:chargeconservation}
 (\frac{\pi}{2}-a_1)=\frac{1}{2}({\pi}-b-a_2) \ .
 \ee
 Using the same arguments we can write down an equation describing
 the conservation of energy
 \be \label{energyconservation}
 4 m \cos(a_1) \cosh(\t) - 2 m\cos(b) = 2 m \cos(a_2) \ ,
 \ee
 where the $-2 m\cos(b)$ term is the boundary energy contribution
 when the field is zero. From (\ref{eq:chargeconservation}) and
 (\ref{energyconservation}) we can determine the rapidity $\theta$
 at which the boundary bound state is formed
 \be
 \t_{n}^{(0,n)} = i \frac{\pi}{k}\left(n+B\right) &\mbox{where}&
 B=\frac{k }{\pi}\ b-\frac{k}{2} \ .
 \ee
 For the above relation the quantisation condition of (\ref{eq:quant1})
 was also used. Now that we have determined where the pole should
 be,
 we need to express it in block notation. It is easy to see that
 since  $(x)$ has a pole in the denominator at $\t = i\frac{\pi}{k}
 x$, we therefore need a block $(n+B)$ and its counterpart $(k-n-B)$
 for the CDD factor of $K_n$.

 Nevertheless the blocks containing the pole should not be the only
 components of the CDD factor. On the contrary we expect the
 number of CDD factors in $K_n$ to increase as $n$ increases towards
 $\frac{k}{2}$ and then to decrease as it approaches $k-1$. Since
 we have $k-1$ particles and the theory is $Z_k$ symmetric we expect
 $K_{k+1}= K_1$. Moreover since the reflection matrix remains the same
 under charge conjugation we expect $K_n=K_{k-n}$.

 Although it is not easy to derive the general formula for
 the CDD factors in the $K_n$ reflection matrix, we can decide on a
 minimal choice for $K_1$. We opt for a CDD factor containing only the
 first pole and its' charge conjugate counterpart. The full reflection
 factor then reads
 \be \label{eq:k10}
 K_1^{(0)}= (1+B)(k-1-B)(1-k) \ .
 \ee
 The superscript denotes the boundary excitation state. In this
 particular case $K_1^{(0)}$ describes the reflection of a soliton
 of charge $Q=1$ from an unexcited boundary. We ignore the poles
 coming from the non-CDD part of $K_1^{(0)}$. We shall address
 this issue at the closing stages of this section. The pole associated
 with the $(1+B)$ factor corresponds to the lowest boundary bound
 state with energy
 \be
E_b^{(1)} = \frac{k m}{2 \pi}\cos(\pi - b -\frac{2 \pi}{k}) \ .
 \ee
 which is formed when a soliton of charge $Q=1$ (which is the particle of the theory) fuses with
 the boundary. The conjugate term $(k-1-B)$ also has a pole but does
 not lie within the physical strip $\Im(\t)\ \epsilon \ (0,\frac{\pi}{2})$. Note that 
 in the classical limit $k\rightarrow \infty$ the reflection factor
 \be
  K_1^{(0)}\rightarrow\frac{i\sinh(\t)+\cos(b)}{i\sinh(\t)-\cos(b)}
  \ee
  in agreement with the identification of the charge $1$ soliton as a particle and
  the classical particle reflection factor (\ref{eq:classicalref}).

\begin{figure}
 \begin{picture}(500,200)(-30,0)
 \put(30,180) {$K_1^{(0)}$}
 \put(30,120) {$K_1^{(1)}$}
 \put(30,60)  {$K_1^{(2)}$}
 \put(30,0)   {$K_1^{(k-1)}$}

 \put(150,180) {$K_2^{(0)}$}
 \put(150,120) {$K_2^{(1)}$}
 \put(150,60)  {$K_2^{(2)}$}
 \put(150,0)   {$K_2^{(k-1)}$}

 \put(230,180) {$K_{k-2}^{(0)}$}
 \put(230,120) {$K_{k-2}^{(1)}$}
 \put(230,60) {$K_{k-2}^{(2)}$}
 \put(230,0) {$K_{k-2}^{(k-1)}$}

 \put(350,180) {$K_{k-1}^{(0)}$}
 \put(350,120) {$K_{k-1}^{(1)}$}
 \put(350,60) {$K_{k-1}^{(2)}$}
 \put(350,0) {$K_{k-1}^{(k-1)}$}

 \put (65,185){\vector(1,0) {80}}
 \put (65,125){\vector(1,0) {80}}
 \put (65,65) {\vector(1,0) {80}}
 \put (65,5)  {\vector(1,0) {80}}

 \put (260,185){\vector(1,0) {80}}
 \put (260,125){\vector(1,0) {80}}
 \put (260,65) {\vector(1,0) {80}}
 \put (260,5)  {\vector(1,0) {80}}

 \put (35,170)  {\vector(0,-1){30}}
 \put (35,110)  {\vector(0,-1){30}}

 \put (155,170) {\vector(0,-1){30}}
 \put (155,110) {\vector(0,-1){30}}

 \put (240,170)  {\vector(0,-1){30}}
 \put (240,110)  {\vector(0,-1){30}}

 \put (240,50) {{\footnotesize .}}
 \put (240,35) {{\footnotesize .}}
 \put (240,20) {{\footnotesize .}}

 \put (360,50) {{\footnotesize .}}
 \put (360,35) {{\footnotesize .}}
 \put (360,20) {{\footnotesize .}}
 \put (360,170)  {\vector(0,-1){30}}
 \put (360,110)  {\vector(0,-1){30}}

 \put (155,50) {{\footnotesize .}}
 \put (155,35) {{\footnotesize .}}
 \put (155,20) {{\footnotesize .}}

 \put (35,50) {{\footnotesize .}}
 \put (35,35) {{\footnotesize .}}
 \put (35,20) {{\footnotesize .}}

 \put (35,50) {{\footnotesize .}}
 \put (35,35) {{\footnotesize .}}
 \put (35,20) {{\footnotesize .}}

 \put (170,185){{\footnotesize \ .  \ .  \ . \ . \ . \ .}}
 \put (170,125){{\footnotesize \ .  \ .  \ . \ . \ . \ .}}
 \put (170,65) {{\footnotesize \ .  \ .  \ . \ . \ . \ .}}
 \put (170,5)  {{\footnotesize \ .  \ .  \ . \ . \ . \ .}}

 \put (80,187) {{\footnotesize Reflection}}
 \put (80,177) {{\footnotesize Bootstrap}}
 \put (-20,160) {{\footnotesize Boundary}}
 \put (-20,150) {{\footnotesize Bootstrap}}
 \end{picture}
 \caption{The Bootstrap Programme}\label{fig:bootprogramme}
 \end{figure}
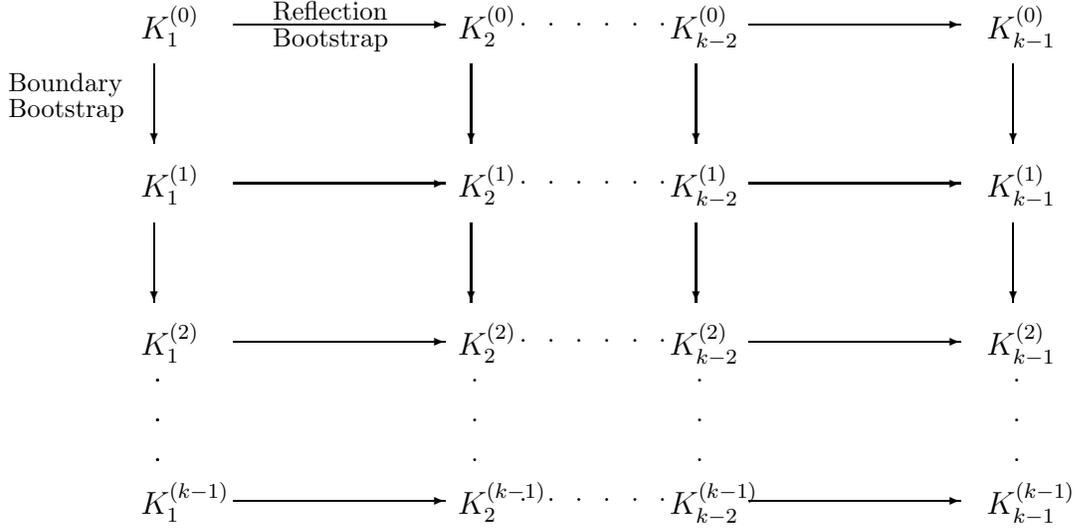

 With $K_1^{(0)}$ as a starting point we can construct the whole
 set of $K_n^{(0)}$ factors by using the reflection bootstrap
 (\ref{eq:reflectionbootstrap}).  In addition, we can apply the
 boundary bootstrap
 \be \label{eq:boundarybootstrap}
 K_b^{(\b)} = S_{ab} (\t_b - \t_a^{\alpha \b})  K_b^{(\alpha)}
 S_{ab} (\t_b + \t_a^{\alpha \b}) \ ,
 \ee
 to each of them to
 obtain the corresponding reflection matrices from the
 excited boundary. In both cases new block factors are generated
 carrying poles which indicate new bound states. If our choice for
 the CDD factor in (\ref{eq:k10}) is correct we expect the bootstrap
  to close, i.e. to end up with a finite spectrum of states which
 is repeated after $k$ steps. Pictorially this can be seen in
 (Fig. \ref{fig:bootprogramme})

 As a starting point we can calculated the reflection factor of a
 charge $Q=2$ soliton bouncing off the unexcited boundary
 \be
 K_2^{(0)}(\t) = K_1^{(0)}(\t-\frac{i \pi}{k})K_1^{(0)}(\t+\frac{i
 \pi}{k})S_{1,1}(2\t) .
 \ee
 Substituting $K_1^{(0)}$ from (\ref{eq:k10}) and $S_{1,1}$ from
 (\ref{eq:smatrix}) we finally get
 \be
 K_2^{(0)} = (2+B)(k-2-B)(B)(k-B)(1-k)(1)(2-k) \ .
 \ee
 This result confirms that the general formula of (\ref{eq:reflectionbase})
 forms a complete set. Henceforth we shall ignore the basic blocks
 of the reflection matrix as they are a consistent set and
 concentrate only on the CDD factors which carry all necessary
 information about boundary bound states. The first four terms  are the
 CDD factor for  $K_2^{(0)}$ which introduce two new poles. An equally
 straightforward calculation is that of $K_4^{(0)}$ which is found to be
 \be
 K_4^{(0)}= \frac{(B+4)(B+2)(k-B+4)(k-B+2)}{
               (-B)(-B+2)(-k+B+2)(-k+B+4)} \ .
 \ee
 A pattern begins to emerge from these simple results. The CDD factor of
 $K_n^{(0)}$ always involves  the block pair $(n+B)(k-n-B)$. This is
 the first consistent appearance of poles in the reflection matrices.
 This is also the complete set of poles that are produced by the bootstrap
 programme. With an arbitrary coupling constant $k$, it is not
 easy to demonstrate the complete mechanism of pole production.
 This can easily be seen by fixing $k$ to an integer value. One
 can then see the whole range of poles produced and how the bootstrap
 miraculously closes.
 \\

 \subsection{The boundary bootstrap}
 \

 In the previous section we proposed a suitable expression for the
 $K_1^{(0)}$ reflection factor and saw how through the reflection
 bootstrap all the $K_n^{(0)}$ factors can be obtained. We now
 turn to the boundary bootstrap to construct the boundary
 reflection factors of solitons bouncing off an excited boundary.
 We shall only examine the boundary bootstrap for the particle
 reflection factor. Through the reflection bootstrap all other
 states can be built (Fig. \ref{fig:bootprogramme}). Once again we are
 facing the problem that with $k$ arbitrary and a charge-varying \sm we
 are not in a position to identify all poles or cancel terms in
 the reflection matrix. We therefore shall attempt to find a
 general formula for the energy difference between adjacent
 boundary bound states. This should be enough to build the whole
 spectrum of states beginning with the energy of the first bound
 state.

 We begin with the boundary bootstrap equation
 (\ref{eq:boundarybootstrap}) for $K_1^{(1)}$ which is
 \be
 K_1^{(1)} = K_1^{(0)} S_{1,1}(\t+ \t_{1}^{(0,1)}) S_{1,1}(\t -
 \t_{1}^{(0,1)}) \ ,
 \ee
 where $\t_{1}^{(0,1)} = i \frac{\pi}{k}\left(1+B\right)$ denotes
 the pole in $K_1^{(0)}$. The product of $S$-matrices in the above
 relation is equal to a shift in their charge. In general the
 following relation holds for any $\psi$.
 \be
 S_{1,1}(\t+ i\psi) S_{1,1}(\t -i \psi) = (2+
 \frac{k}{\pi}\psi)(2- \frac{k}{\pi}\psi) \ .
 \ee
 Putting everything together yields
 \be
 K_1^{(1)} = (1+B)(k-1-B) (3+B)(1-B) \ .
 \ee
 Once again we have only used the CDD factors and ignored terms
 not involving the boundary constant. We see through this
 procedure a new pole appears from the block $(3+B)$ corresponding
 to the absorption of a charge $Q=1$ particle into the charge $Q_b=1$
 boundary.  This will be used as an input to find a new pole in $K_1^{(2)}$
 \be
 K_1^{(2)} = (k-1-B) (3+B)(1-B) (5+B) \ ,
 \ee
 where a cancellation has already taken place. The new pole comes
 from the block factor $(5+B)$ which will be used in the next
 step. The expression for $K_1^{(3)}$ is
 \be
 K_1^{(3)} = (k-1-B) (1-B) (5+B) (7+B) \ ,
 \ee
 where new factor $(-3-B)$ has cancelled with the block factor $(3+B)$
 and the factor $(7+B)$ has introduced a new pole. This procedure
 will not continue indefinitely. After $k$ steps we expect to
 return to original form of $K_1^{(0)}$. However it is clear that
 the $K_1^{(n)}$ reflection matrix will have a pole indicated by
 the block factor $(2n+1+B)$ at $ \psi_n =  \frac{\pi}{k}(2n+1+B)$.
 Having a general formula about the $n$-th pole allows us to write
 down a recursive relation for the energies of the bound states.
 We begin with the energy of the first bound state. The difference
 between the first excited boundary state and the non-excited
 boundary is
 \be
 E_1 - E_0 = A \cos (\psi_0) \ .
 \ee
 The right hand side is equal to the mass of the incoming particle
 that binds with the boundary at a fixed angle $\t = i \psi_0$.
 The parameter A is related to the mass of the particle
 and will be determined from the comparison with the semi-classical
 results. The formula may be used recursively to finally yield
 \be\label{eq:energydifference2}
 E_{n+1}-E_n = A \cos(\psi_n) = A \cos( \frac{\pi}{k}(2n+1+B)) \ .
 \ee
 This formula should be compared with the one derived from the
 semi-classical approach (\ref{eq:energydifference}). Both
 describe the exact same energy gaps between two bound states.
 The arbitrary parameter $A$ in (\ref{eq:energydifference2}) can be
 read directly from (\ref{eq:energydifference})
 \be
 A =\frac{k m}{\pi}\cos\left(\frac{\pi}{2}-\frac{\pi}{k}\right)= \frac{k m}{\pi}\sin\left(\frac{\pi}{k}\right )
  \ee
 which is equal to the energy of the particle. In the limit that $k\rightarrow
 \infty$ (the classical limit) the soliton becomes infinitesimally
 small and the boundary-bound states spectrum become continuous.

 We conclude this section with a brief discussion about the poles
 appearing in the reflection matrix. We expect the non-CDD poles in
 (\ref{eq:reflectionbase}) to be explained in terms of a on-shell
 triangle diagram as is the case for the $a_n^{(1)}$ affine Toda
 theory. A full discussion can be found in \cite{Delius:1999cs}.
 The remaining poles appear in the CDD factors and are the ones
 associated with the boundary-bound states.
 The full set of poles generated from the bootstrap programme come
 from the general blocks $(n+B)$ and
 $(k-n-B)$ which have poles at
 \be
  \psi_n= \frac{\pi}{k}n -\frac{\pi}{2} + b   &\mbox{and}&
  \psi_n'= \pi - \left(\frac{\pi}{k}n -\frac{\pi}{2} + b\right)
 \ee
 From the forms above we can see that if $\t_n$ lies in the
 physical strip $(0,\frac{\pi}{2})$ then $\t_n'$ does not, and vice versa.
 Assuming that the pole is at $\t_n$ then it lies in the physical
 strip if
 \be
 \frac{\pi}{2} > \psi_n > 0 &\Leftrightarrow& 1> \frac{n}{k} + \frac{b}{\pi} >
 \frac{1}{2}
 \ee
 As soon the above condition is no longer true then the pole appears in
 the conjugate block at $\psi_n'$. This however does not alter any
 of our results. From (\ref{eq:energydifference2}) we can see that
 $\psi_n$ or $\psi_n'$ appear in the argument of the cosine so
 both correspond to the same absolute energy difference between two states.
 \\

 \section{Discussion}
 \

 We used semi-classical method to obtain the boundary spectrum of the CSG model. Having incorporated the unitary and crossing constraints, we found that we needed to adjust our simplest guess for the reflection factor of a particle off an unexcited boundary by a CDD factor, in order that the reflection factor in question exhibited a pole corresponding to the lowest bound state of the boundary.
 Reflection factors for higher charge solitons, and for particles and solitons scattering off excited boundaries were obtained by iterating the reflection bootstrap and boundary bootstrap respectively. The rest of the excited boundary spectrum could be deduced by analysing the poles in these higher reflection factors, and, perhaps remarkably, it coincides with that obtained by semi-classical methods. The reflection and boundary bootstrap both
 close because the minimal choice for $K_n$ (\ref{eq:reflectionbase})
 is consistent with our \sm and because the introduced blocks are CDD
 factors.

 One final point to be made is that although we have used for our
 comparison equations (\ref{eq:quant1}) and
 (\ref{eq:energydifference}) which correspond to the first
 approximation in the semi-classical approach, we can extend our
 results to agree with the first order corrections by simply
 substituting everywhere $k=k_r$ and $b=b_r$. The redefinition of
 both parameters change nothing in the bootstrap approach. If we believe, as 
 seems to be the case for the bulk coupling constant, that the boundary coupling
 constant receives no further corrections at higher orders in perturbation theory, then
 the formulae containing $k_r$ and $b_r$ would be exact. 

\bibliographystyle{unsrt}

\begin{thebibliography}{10}

\bibitem{Lund:1976ze}
Fernando Lund and Tullio Regge.
\newblock Unified approach to strings and vortices with soliton solutions.
\newblock {\em Phys. Rev.}, D14:1524, 1976.

\bibitem{Pohlmeyer:1975nb}
K.~Pohlmeyer.
\newblock Integrable hamiltonian systems and interactions through quadratic
  constraints.
\newblock {\em Commun. Math. Phys.}, 46:207, 1976.

\bibitem{Fernandez-Pousa:1997hi}
Carlos~R. Fernandez-Pousa, Manuel~V. Gallas, Timothy~J. Hollowood,
and J.~Luis
  Miramontes.
\newblock The symmetric space and homogeneous sine-gordon theories.
\newblock {\em Nucl. Phys.}, B484:609--630, 1997.

\bibitem{Fernandez-Pousa:1998iu}
Carlos~R. Fernandez-Pousa and J.~Luis Miramontes.
\newblock Semi-classical spectrum of the homogeneous sine-gordon theories.
\newblock {\em Nucl. Phys.}, B518:745--769, 1998.

\bibitem{Castro-Alvaredo:1999em}
O.~A. Castro-Alvaredo, A.~Fring, C.~Korff, and J.~L. Miramontes.
\newblock Thermodynamic bethe ansatz of the homogeneous sine-gordon models.
\newblock {\em Nucl. Phys.}, B575:535--560, 2000.

\bibitem{Dorey:2002sc}
  P.~Dorey and J.~L.~Miramontes,
\newblock Mass scales and crossover phenomena in the homogeneous sine-Gordon models
 \newblock{\em Nucl.  Phys.}  B697:  405-461,2004.
  
\bibitem{Getmanov:1977hk}
B.~S. Getmanov.
\newblock New lorentz invariant systems with exact multi - soliton solutions.
  (in russian).
\newblock {\em Pisma Zh. Eksp. Teor. Fiz.}, 25:132--136, 1977.

\bibitem{Tzamtzis:2002}
Peter Bowcock and Georgios Tzamtzis.
\newblock The complex sine-Gordon model on a half-line
\newblock hep-th/0203139

\bibitem{Ghoshal:1994tm}
Subir Ghoshal and Alexander~B. Zamolodchikov.
\newblock Boundary s matrix and boundary state in two-dimensional integrable
  quantum field theory.
\newblock {\em Int. J. Mod. Phys.}, A9:3841--3886, 1994.

\bibitem{deVega:1981ka}
H.~J. de~Vega and J.~M. Maillet.
\newblock Renormalization character and quantum s matrix for a classically
  integrable theory.
\newblock {\em Phys. Lett.}, B101:302, 1981.

\bibitem{deVega:1983sh}
H.~J. de~Vega and J.~M. Maillet.
\newblock Semiclassical quantization of the complex sine-gordon field theory.
\newblock {\em Phys. Rev.}, D28:1441, 1983.

\bibitem{Dashen:1975hd}
Roger~F. Dashen, Brosl Hasslacher, and Andre Neveu.
\newblock The particle spectrum in model field theories from semiclassical
  functional integral techniques.
\newblock {\em Phys. Rev.}, D11:3424, 1975.

\bibitem{Dashen:1974ci}
Roger~F. Dashen, Brosl Hasslacher, and Andre Neveu.
\newblock Nonperturbative methods and extended hadron models in field theory.
  1. semiclassical functional methods.
\newblock {\em Phys. Rev.}, D10:4114, 1974.

\bibitem{Bonneau:1983hx}
Guy Bonneau.
\newblock Renormalizability and nonproduction in complex sine-gordon model.
\newblock {\em Phys. Lett.}, B133:341, 1983.

\bibitem{Dorey:1995mg}
Nicholas Dorey and Timothy~J. Hollowood.
\newblock Quantum scattering of charged solitons in the complex sine- gordon
  model.
\newblock {\em Nucl. Phys.}, B440:215--236, 1995.

\bibitem{Bakas:1994xh}
Ioannis Bakas.
\newblock Conservation laws and geometry of perturbed coset models.
\newblock {\em Int. J. Mod. Phys.}, A9:3443--3472, 1994.

\bibitem{Ventura:1976vi}
Ivan Ventura and Gil~C. Marques.
\newblock The bohr-sommerfeld quantization rule and charge quantization in
  field theory.
\newblock {\em Phys. Lett.}, B64:43, 1976.

\bibitem{Montonen:1976yk}
C.~Montonen.
\newblock On solitons with an abelian charge in scalar field theories. 1.
  classical theory and bohr-sommerfeld quantization.
\newblock {\em Nucl. Phys.}, B112:349, 1976.

\bibitem{deVega:1994pm}
H.~J. de~Vega, J.~Ramirez~Mittelbrunn, M.~Ramon~Medrano, and
N.~Sanchez.
\newblock The general solution of the 2-d sigma model stringy black hole and
  the massless complex sine-gordon model.
\newblock {\em Phys. Lett.}, B323:133--138, 1994.

\bibitem{Penati:1995bs}
S.~Penati and D.~Zanon.
\newblock Quantum integrability in two-dimensional systems with boundary.
\newblock {\em Phys. Lett.}, B358:63--72, 1995.

\bibitem{Cherednik:1984vs}
I.~V. Cherednik.
\newblock Factorizing particles on a half line and root systems.
\newblock {\em Theor. Math. Phys.}, 61:977--983, 1984.

\bibitem{Fring:1994mp}
Andreas Fring and Roland Koberle.
\newblock Factorized scattering in the presence of reflecting boundaries.
\newblock {\em Nucl. Phys.}, B421:159--172, 1994.

\bibitem{Delius:1999cs}
Gustav~W. Delius and Georg~M. Gandenberger.
\newblock Particle reflection amplitudes in a(n)(1) toda field theories.
\newblock {\em Nucl. Phys.}, B554:325--364, 1999.

\bibitem{Corrigan:1999fp}
  E.~Corrigan and G.~W.~Delius,
\newblock Boundary breathers in the sinh-Gordon model
\newblock{\em { J. Phys. A}} {\bf 32} (1999) 8601.

\end{thebibliography}

\end{document}